\documentclass[trackchanges,preprint2]{aastex6}
\turnoffedit

\begin{document}


\title{Intensity Modulation of Fluorescent Line by a Finite Light Speed Effect in Accretion-Powered X-ray Pulsars}


\author{Yuki Yoshida\altaffilmark{1,2}, Shunji Kitamoto\altaffilmark{1,2}}
\and
\author{ Akio Hoshino\altaffilmark{1,2}}

\altaffiltext{1}{Department of Physics, College of Science, Rikkyo University, 3-34-1 Nishi-Ikebukuro, Toshima, Tokyo 171-8501, Japan}
\altaffiltext{2}{Research Center for Measurement in Advanced Science, Rikkyo University, 3-34-1 Nishi-Ikebukuro, Toshima, Tokyo 171-8501, Japan}
\email{yy@rikkyo.ac.jp}

\begin{abstract}
The X-ray line diagnostic method is a powerful tool for an investigation of plasma around accretion-powered X-ray pulsars. 
We point out an apparent intensity modulation of emission lines with their rotation period of neutron stars, due to the finite speed of light (we call this effect "finite light speed effect"), 
if the line emission mechanism is a kind of reprocessing, such as fluorescence or recombination after ionization by X-ray irradiation from pulsars.
The modulation amplitude is determined by the size of the emission region, 
in competition with the smearing effect by the light crossing time in the emission region.
This is efficient if the size of the emission region is roughly comparable to that of  the rotation period multiplied by the speed of  light.
We apply this  effect to a symbiotic X-ray pulsar, GX\,1+4, a spin modulation of the intense iron line of which has been reported.
The finite light speed effect can explain the observed intensity modulation, if its fluorescent region has a size of $\sim 10^{12}$cm.
 
\end{abstract}




\section{Introduction}
Many of the accretion powered X-ray pulsars exhibit prominent iron emission lines from neutral atoms or ions. 
A detailed investigation of these emission lines is useful for diagnostics of their emission mechanism. 
Identification of emission lines (element and its ionization state) and their observed variability in the line parameters with time and with source luminosity provide important information on the line emission mechanism and their emission regions.
Several possible line-emission regions have been considered  such as  inhomogeneous and clumpy stellar wind from the companion star \citep{Sako2002,Wojdowski2003}, surface or photosphere of the companions star, and the accreting matter around the Alfv\'en shell \citep{Basko1980, Vrtilek2005, Kohmura2001}. 

Bright  Neon and Oxygen emission lines (OVII) from 4U\,1626-67 \edit3{were} first reported by \citet{Angelini1995}, which is a low mass X-ray binary pulsar with 
 7.7~s spin-period.
\edit3{Flux modulation of the Oxygen line, by a factor of up to 4, was reported by \citet{Beri2015},} supporting the warped accretion-disk origin  \citep{Schulz2001}.
Hercules\,X-1 is another X-ray pulser with \edit3{a} spin period of 2.3~s.
Its iron line shows an intensity modulation  with its spin period \citep{Vasco2013, Zane2004, Choi1994}.
\citet{Zane2004} and \citet{Choi1994} reported that the intensity modulation of the emission line is roughly represented by a sinusoidal function, 
the bottom phase of which is corresponding to the peak phase of the pulse shape of the hard X-ray continuum.
But, \citet{Vasco2013} reported disappearance of the iron line flux around the peak phase of the pulse shape of the hard X-ray continuum, 
and speculated a  hollow cone geometry for the accretion column.
For these fast rotators, \edit3{the emission region is expected to be located either near the vicinity of the neutron star or at the inner part of the accretion disk.}

The Fe fluorescent line is notable in the spectrum of GX\,301-2, the spin period of which is $\sim$700~s, but the question of the fluorescent region is in a vigorous debate.
\citet{Endo2002} analyzed the line shape of the intense Fe emission line  from GX\,301-2 and they found a significantly broader width of the line than that estimated based on the terminal velocity of the stellar wind.
\edit3{They concluded that the Fe lines originate within 10$^{10}$~cm of the continuum emission source, i.e. inside of the accretion radius.}
\citet{Furst2011} reported  a significantly small pulse fraction of the Fe emission lines but a strong pulse-to-pulse variation in its intensity.  
They investigated a correlation with the continuum flux and the Fe emission line flux, and showed that the Fe emission region is not far from the X-ray source, which is consistent with the result by \citet{Endo2002}.
On the other hand, \citet{Suchy2012} reported  results by {\it Suzaku} observation and 
they argued that the line flux did not change significantly throughout the pulse phase, and concluded that the Fe fluorescence region was greater than
$\sim$ 700 lt-s ($\sim 2\times 10^{13}$ cm) from the neutron star, 
although  \edit3{there is a tendency toward higher fluxes around phase 0.6--1.0,} where the pulse profile of the continuum X-rays shows the second peak, with an amplitude of $\sim$ 10~\%.

The origin of emission lines in GX\,1+4, which is a symbiotic X-ray pulsar with $\sim 150$~s spin period,  was discussed by \citet{Kotani1999} and \cite{Yoshida2017} (hereafter Paper I).
\edit3{Based on the ionization state of the iron atoms, the absorption column density, and estimated X-ray luminosity derived from an {\it ASCA} observation of GX\,1+4, 
\citet{Kotani1999} suggested that the line emitting region in GX\,1+4 consisted of low ionization matter extending to greater than $10^{12}$~cm from the neutron star. 
This radius was estimated with the assumption that the matter was distributed homogeneously.}
The size of emitting region of $r>3.4\times10^{12}$~cm in radius was estimated with {\it Suzaku} observation adopting the similar method.
It was also suggested that the size of fluorescent region can be reduced to $\sim10^{11}$~cm by introducing inhomogeneity of matter (see Paper I). 
Paper I also reported that the iron emission line shows its intensity modulation with an amplitude of $7\pm1$~\%, according to the pulse phase.

In this paper, we point out an importance of the following  two effects in  the discussion of the observed time variation of the line flux. 
If the fluorescent lines are emitted from such a large region, the observed time variation should be smeared with the light crossing time of the region, 
which is roughly 100~s, for the size of 3.4$\times$10$^{12}$~cm. 
On the other hand, we should take into account  an apparent  intensity modulation of the fluorescent lines by an effect of the finite light speed, as described in the next section.

\section{Finite Light Speed Effect}
We are now considering fluorescent lines from matter exposed to X-rays from an X-ray pulsar,
the emission profile of which  is not spherically symmetric.
Therefore, the fluorescent region, which is illuminated by X-rays from the X-ray pulsar, changes according to the rotation of the neutron star. 
As a consequence, the fluorescent region appears to show an apparent movement to the observer. 
The apparent movement affects the observed intensity per unit time of the fluorescent lines, since the speed of light is finite, 
even if the circumstellar matter around the neutron star is homogenous and uniform and the intensity of X-rays from the neutron star stays constant.
If the fluorescent region is going away from an observer with a velocity of $v_{\rm los}$,  along the line of sight, emitted photons at the fluorescent region in 1~s is observed in $(1+v_{\rm los}/c)$~s by the observer.   
Therefore, the intensity of the fluorescent line observed in 1~s decreases with a factor of  $(1+v_{\rm los}/c)^{-1}$.
On the other hand,  the intensity observed in 1~s increases by a factor of  $(1-v_{\rm los}/c)^{-1}$ at the moment when the emission region is approaching to the observer.
This phenomenon is similar to the Doppler boosting  but the wave length of  fluorescent lines does not change.
Because the gas in the fluorescent matter is not moving but the region of the fluorescent is changing.
We call this the finite light speed effect.

As mentioned in the previous section, the fluorescent region of GX~1+4 is thought to extend greater than 10$^{12}$~cm.  
Since the pulse period of GX~1+4 is $\sim$150~s, the speed of the change of the fluorescent region is comparable to the speed of light.
Therefore, a significant effect of this finite light speed effect can be expected.

\subsection{Model Calculation and Results}
\begin{figure}
\begin{center}
  \includegraphics[width=75mm]{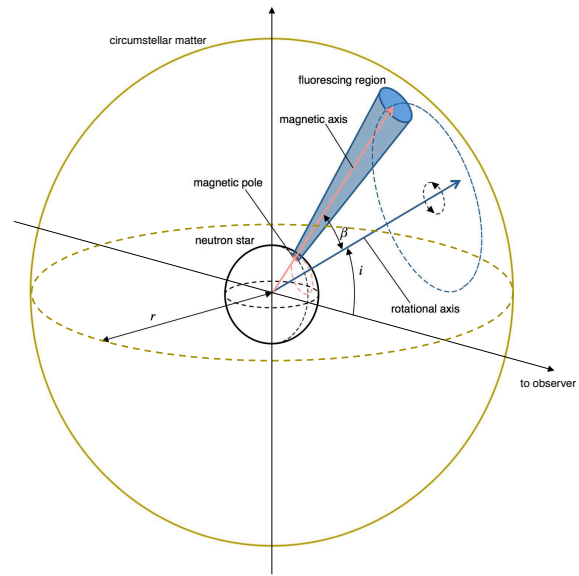}
\end{center}
\caption{Geometry used in the Monte Carlo simulation.
The neutron star is located at the center of the circumstellar matter with the radius of $r$, which is represented by the yellow largest circle. 
Blue and red arrows indicate the rotational axis and the magnetic axis, respectively. 
The blue cone, whose bottom is a magnetic pole, shows the fluorescing region at a certain phase. 
Red and blue circles with dashed-lines indicate the trajectories of the magnetic pole and magnetic axis, respectively. 
The size of the neutron star is enlarged to make it easier to see.}
\label{fig:geometry}
\end{figure}

We demonstrated the finite light speed effect using Monte Carlo simulation.
Figure~\ref{fig:geometry} shows a schematic drawing of a configuration of the simulation.
In the simulation, a neutron star is located at the center of  spherically uniform and homogeneous matter.
We assume that X-rays are emitted at magnetic poles and the magnetic axis is assumed to be tilted with an angle of $\beta$ from the rotational axis.
The inclination angle of the rotational axis is denoted by an angle $i$ between the rotation angle and the line of sight.    
Intensity of the photons from a magnetic pole is assumed to be 
proportional to $\cos{\theta}$, where $\theta$ is the angle of direction of each photon from the magnetic axis.
Although the illuminated region is indicated by the conic geometry for simplicity, the X-rays from the pulsar are emitted with the above given intensity distribution from the pole in our calculation.
\edit3{Based on the optical depth of the absorbing circumstellar material, the X-ray photons will be absorbed with an exponential probability.
Once absorption occurs, a fluorescent photon is immediately emitted that can be detected by the observer.}
We calculate the propagation time of individual photons from the magnetic pole to the observer. 
By assuming a constant emission rate of the photons from the magnetic pole,  we generate an expected light curve  by  enumerating the arrival time of photons. 

\begin{figure}
\begin{center}
  \includegraphics[width=75mm]{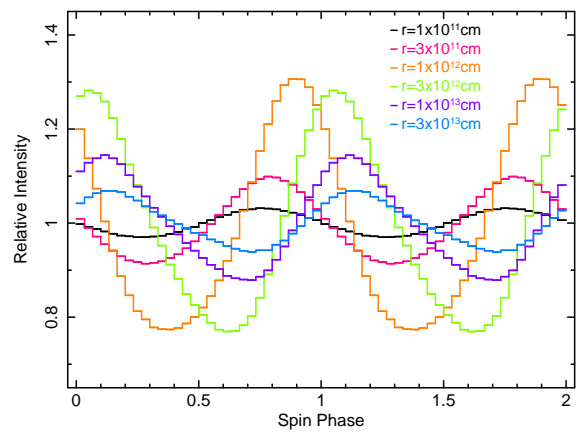}
\end{center}
\caption{Examples of simulated folded light curves of the fluorescent lines as a function of the spin phase of the neutron star.
The light curves calculated by Monte Carlo simulation were folded by the spin period of the neutron star by fixing the time origin (epoch) to be a time when a direct photon from the magnetic pole was detected by the observer.  
We  assumed  a common inclination angle of the rotation axis $i=90$~deg, and angle between rotational axis and magnetic axis $\beta$=45~deg.
Six different radii of the circumstellar matter were assumed, as $r=1\times10^{11}$,  $3\times10^{11}$, $1\times10^{12}$, $3\times10^{12}$, $1\times10^{13}$ and $3\times10^{13}$~cm, which are  indicated by different colors.
The obtained folded light curves were normalized by dividing by the average fluorescent intensity.}
\label{fig:eflcEample}
\end{figure}

We calculated six cases of the radius of the circumstellar matter, $r=1\times10^{11}$, $3\times10^{11}$, $1\times10^{12}$, $3\times10^{12}$, $1\times10^{13}$ and $3\times10^{13}$~cm,  with a spin period $P=150$~s,   $\beta=45$~deg, and $i=90$~deg.
We folded the six generated light curves  on the spin period  by assuming the time origin to be a time when a direct photon from a magnetic pole was detected by the observer.
Intensities of the folded light curves were normalized as the average intensity to be unity and   plotted as  a function of the spin phase  in  Figure~\ref{fig:eflcEample}.
We can see the intensity modulation by the finite light speed effect.
The amplitude of the intensity modulation  becomes larger with increasing radius of the circumstellar matter up to $1\times10^{12}$~cm, 
while, in the cases of $r\ge3\times10^{12}$~cm, the amplitude  gradually becomes smaller than that of $1\times10^{12}$~cm as the radius becomes larger.
We can  also see  phase shifts as a change of the radius of the circumstellar matter, $r$, where a larger phase delay is seen in the  larger radius.

\begin{figure}
\begin{center}
  \includegraphics[width=75mm]{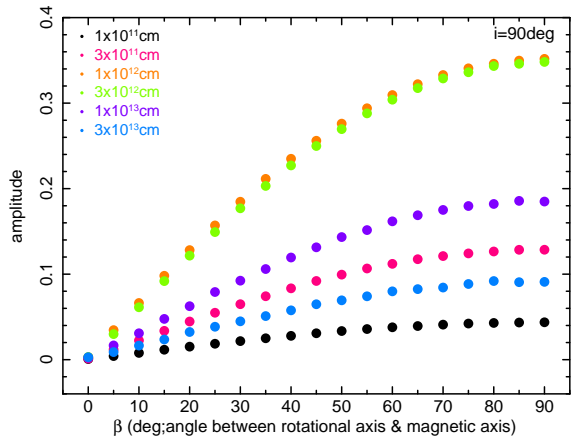}
\end{center}
\caption{Relation between the amplitude of intensity modulation of fluorescent lines and the angle of $\beta$,  assuming the inclination angle of the rotation axis $i=90$~deg.
Six radii in Figure~\ref{fig:eflcEample} were calculated.
Radii of the circumstellar matter are indicated by different colors.}
\label{fig:relbaExample}
\end{figure}

In Figure~\ref{fig:relbaExample}, the modulation amplitude as a function of $\beta$ is plotted for the six cases of the radius of the matter for $i=90.0$~deg.
We can recognize that the modulation amplitude  becomes larger with increasing the angle of $\beta$ from 0-deg to 90-deg.
The dependence on the radius of matter is again found as the same to that shown in  Figure~\ref{fig:eflcEample}, i.e. the amplitude is the largest at the radius of $1\times10^{12}$~cm.

If $i=90.0$~deg, the velocity of the apparent movement of the center of fluorescent region along the line of sight, $v_{\rm los}$, is calculated as $v_{\rm los} \sim \frac {2\pi (r/2) \sin{\beta}\sin({2\pi\phi})}{P}$, for an optically thin case, where $\phi$\,(=0.0--1.0) is the phase of the rotation of fluorescent region from the origin defined as the direction of the observer.
Therefore,  $v_{\rm los}$ becomes large at large  $\beta$ and at large $r$, 
and hence  the modulation amplitude of line intensity by the finite light speed effect becomes larger.
The delay of the maximum phase is also understood by a long light travel time in the large size of the matter.

At a radius of $1\times10^{13}$~cm,  the light crossing time becomes $\sim 300$~s, which is roughly two times larger than the spin period.
Therefore, the intensity modulation by the finite light speed effect is smeared with the light crossing time.
This smearing decreases the amplitude of the modulation.
The competition, between the finite light speed effect and the smearing effect,  determines the amplitude of the modulation and causes  phase-shifts and waveform-variations as we can see in Figure~\ref{fig:eflcEample}.

\subsection{Application to GX\,1+4}
We used the simulation framework described in previous section to mimic intensity modulation of the fluorescent line observed in the symbiotic X-ray pulsar GX\,1+4.
Using {\it Suzaku} observation, Paper I found that iron K$_{\alpha}$ emission line shows the intensity modulation according to the pulse period, with an amplitude of $7\pm1$~\%, peaking at around 0.7--1.1 in phase.

We calculated the arrival time at the observer for each photon considering six cases for a radius of the circumstellar matter,
$r=1\times10^{11}$, $3\times10^{11}$, $1\times10^{12}$, $3\times10^{12}$, $1\times10^{13}$, and $3\times10^{13}$~cm, and the spin period, $P=$~150~s.
The mean free path of the circumstellar matter was fixed at the optical depth of the neutral iron K-edge,  derived in Paper I with {\it Suzaku} observation.
The simulation was conducted for different angle of $i$ and $\beta$ in a range between 0~deg and 90~deg with a 5~deg step, respectively.

\begin{figure*}
\begin{center}
  \includegraphics[width=55mm]{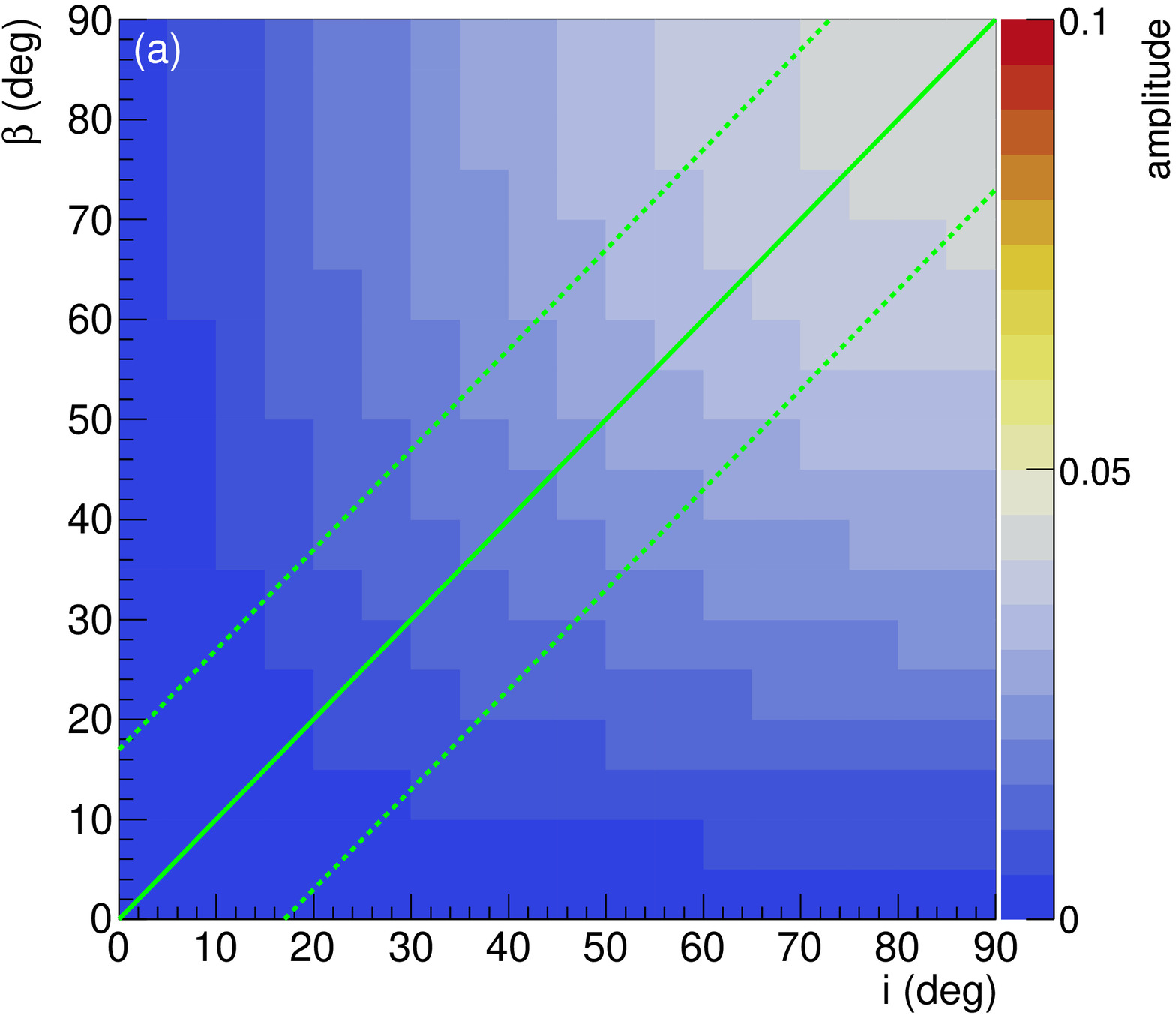}
  \hspace{0.1em}
  \includegraphics[width=55mm]{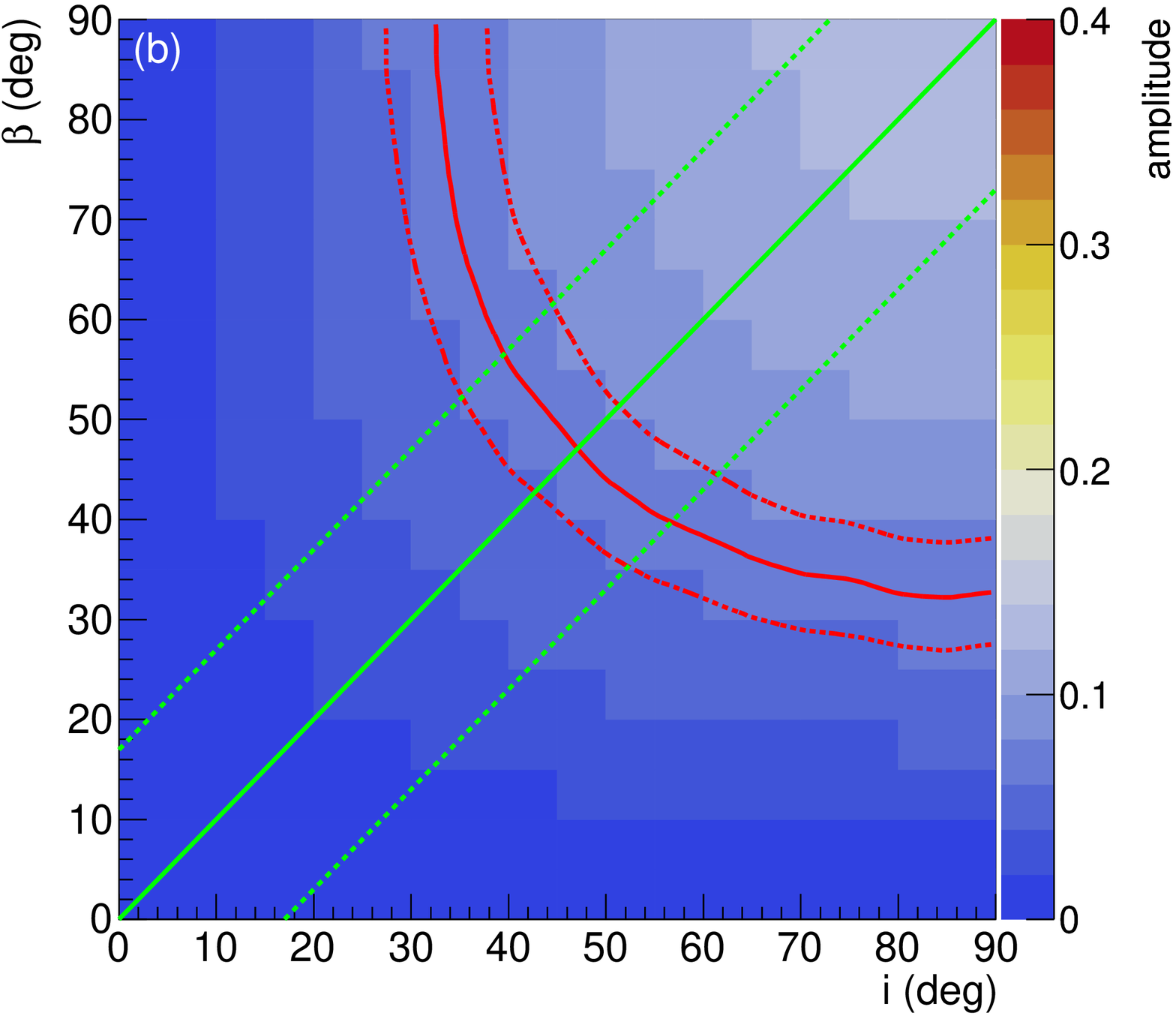}   
  \hspace{0.1em}
  \includegraphics[width=55mm]{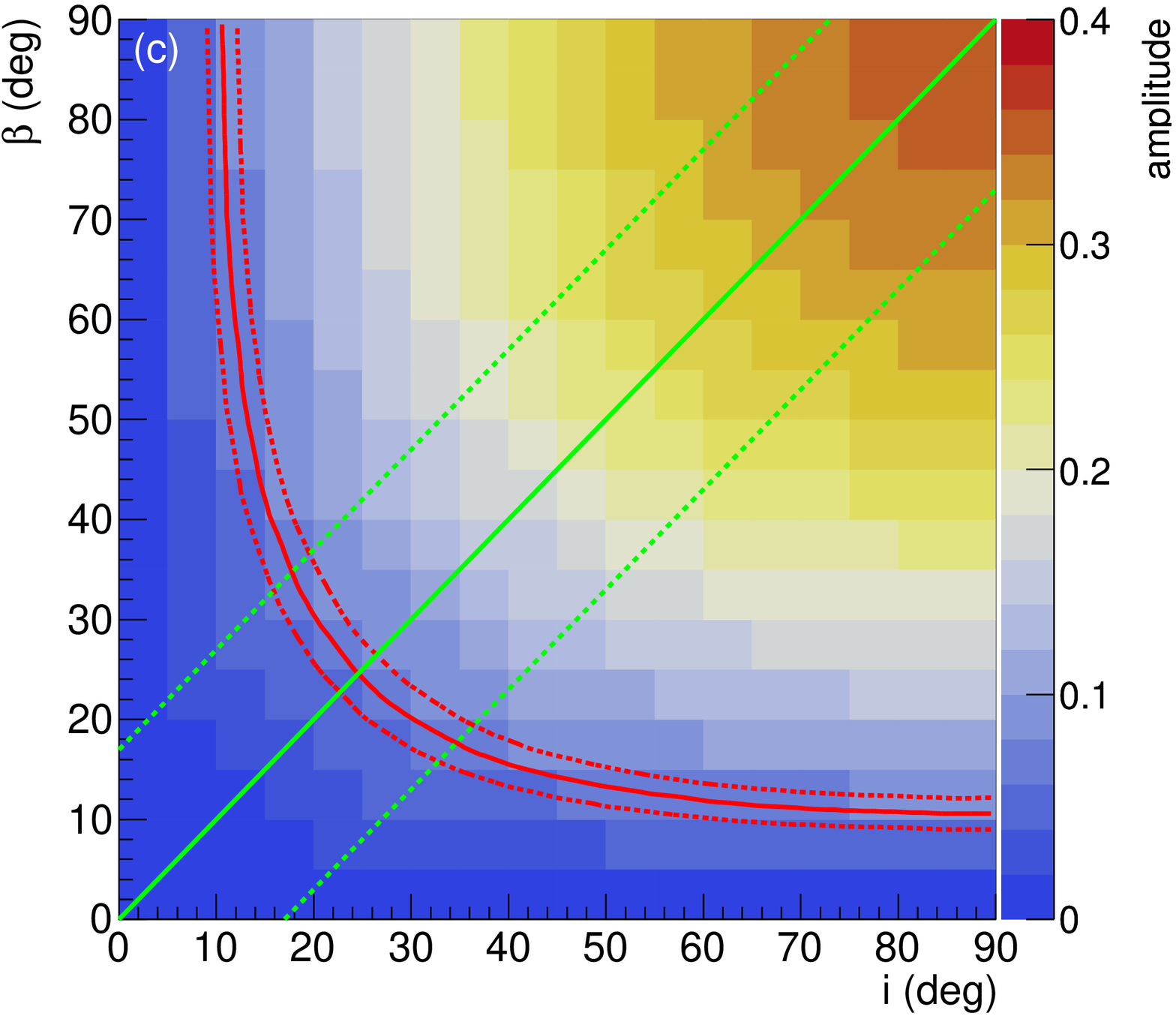} \\
   \vspace{0.5em}
  \includegraphics[width=55mm]{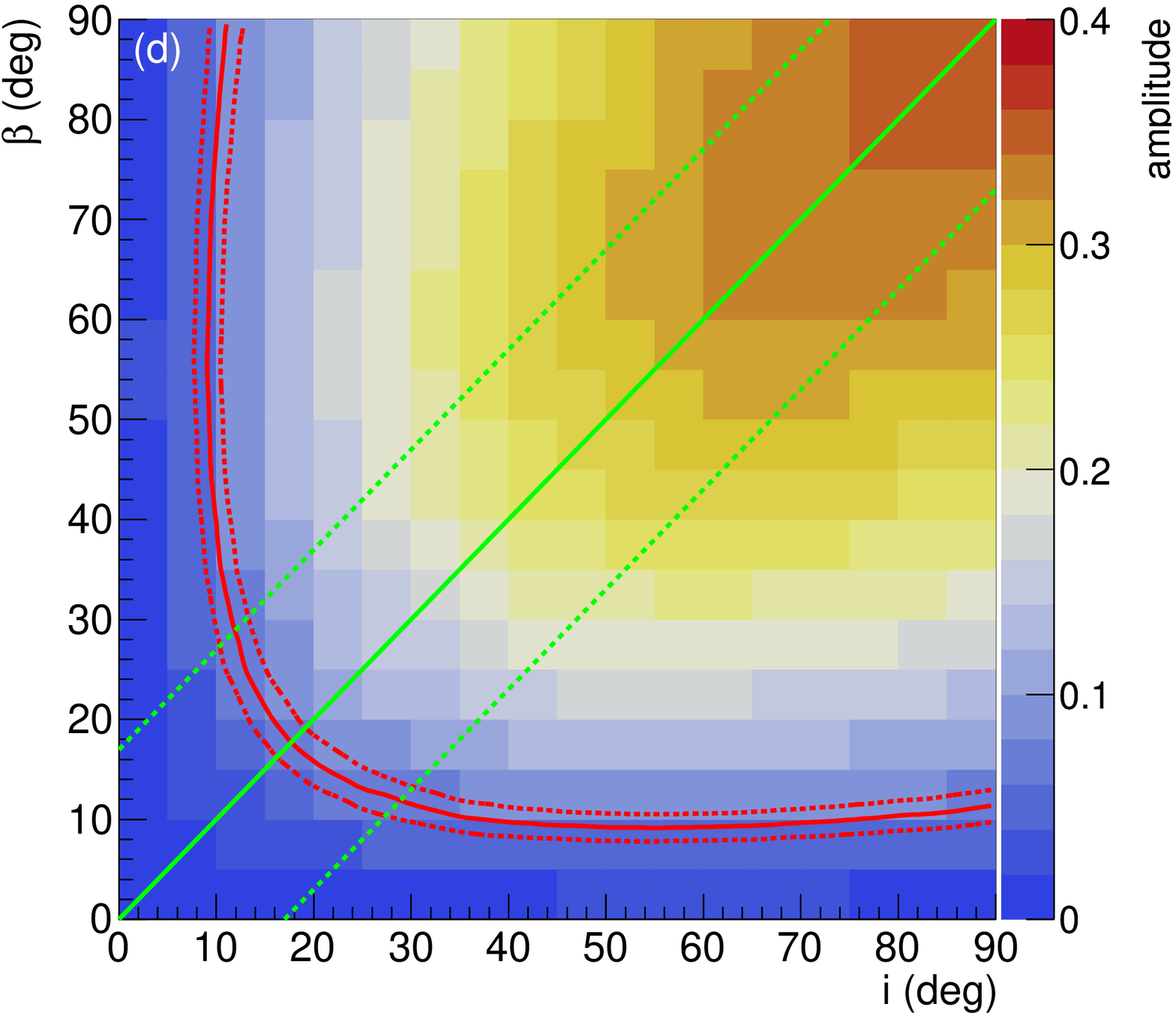} 
  \hspace{0.1em}
  \includegraphics[width=55mm]{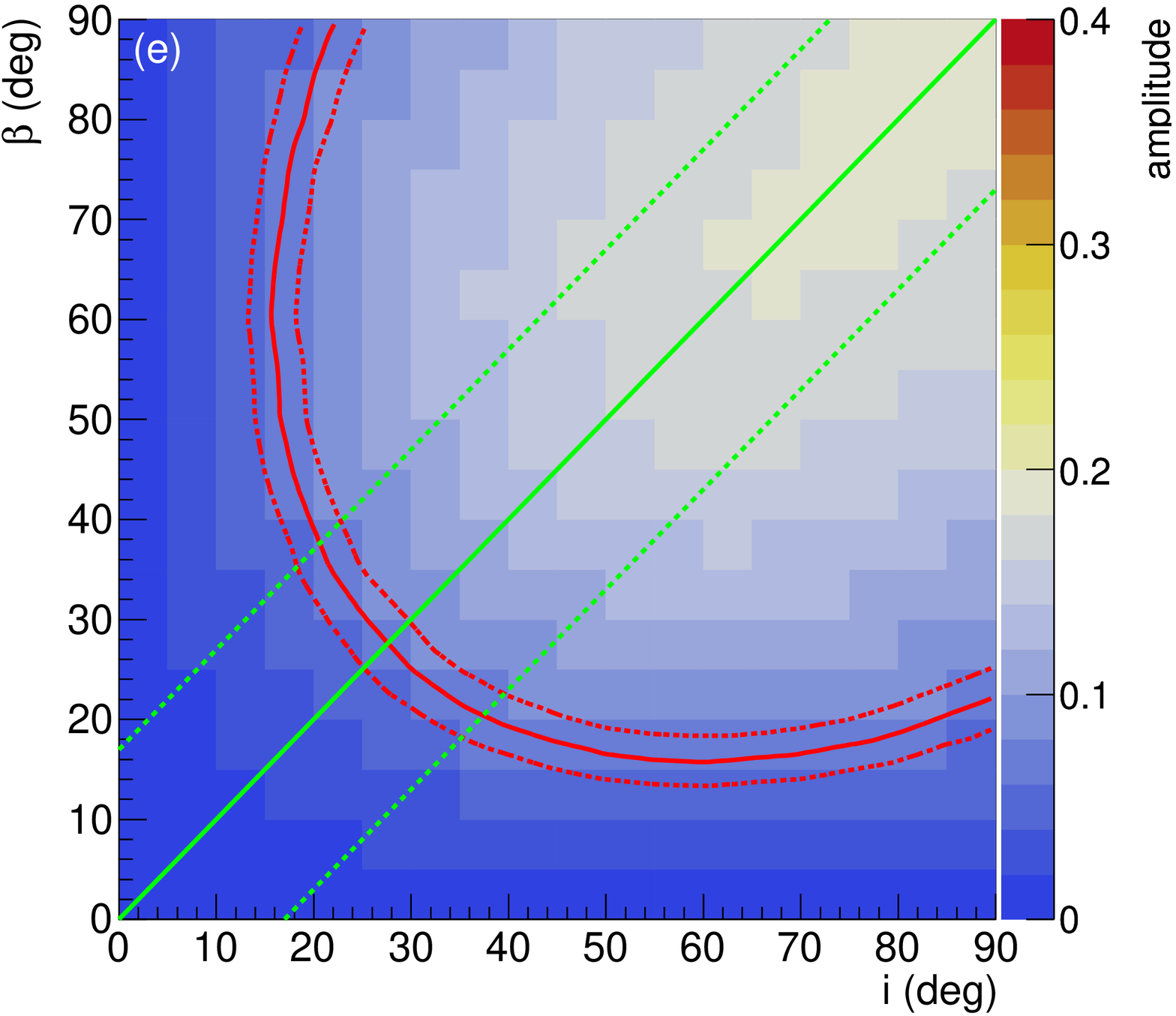}
  \hspace{0.1em}
  \includegraphics[width=55mm]{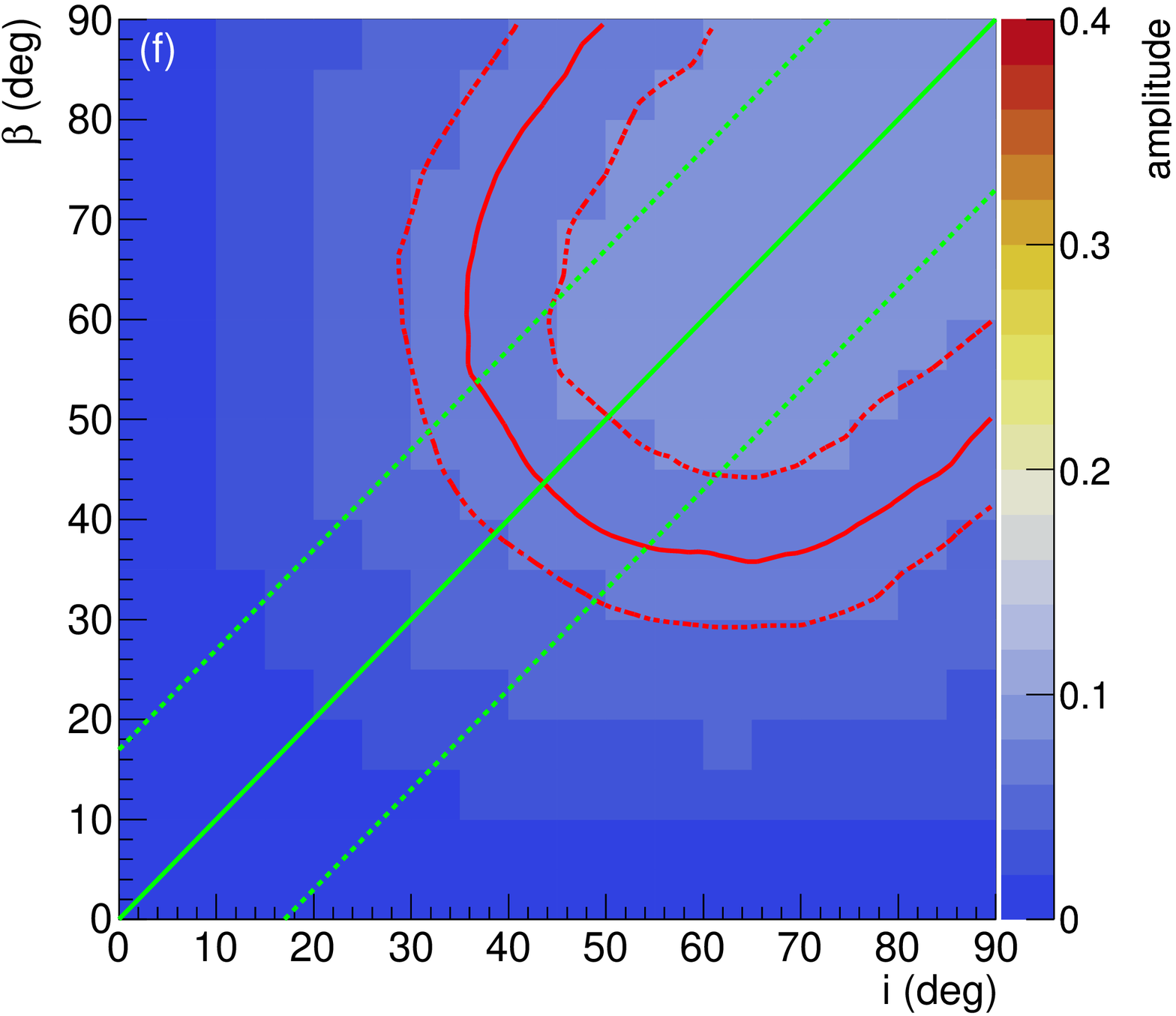}
\end{center}
\caption{The distribution of calculated amplitude of intensity-modulation of fluorescent line on a two-dimensional map of  $i$ and $\beta$.
Six  radii of the circumstellar matter are assumed; (a) $r=1\times10^{11}$~cm, (b) $r=3\times10^{11}$~cm, (c) $r=1\times10^{12}$~cm, (d) $r=3\times10^{12}$~cm, (e) $r=1\times10^{13}$~cm, (f) $r=3\times10^{13}$~cm.
The red-solid and the red-dotted lines indicate the amplitude level of 7~\% and the acceptable regions derived by its errors, respectively.
Green-solid lines  indicate the position of  $i=\beta$. An acceptable range from the pulse profile ($i-17^{\circ} \le \beta \le i+17^{\circ}$) is indicated by green-dotted lines. 
}
\label{fig:ampdist}
\end{figure*}

Figure~\ref{fig:ampdist} shows the calculated amplitude of intensity modulation of fluorescent line on two dimensional maps of the angles,  $i$ and $\beta$ for each radius.
The calculated amplitude of the time variation for $r=1\times10^{11}$~cm is less than 7~\% in any combination of the angles,
while in cases of $r\ge3\times10^{11}$~cm, a modulation amplitude of  7~\% level can be seen in some combinations of $i$ and $\beta$.
The red-solid lines indicate the position of the amplitude of 7~\%  
and the red-dotted lines indicate the acceptable region by uncertainty (1~\%) of the observed amplitude, which were obtained from {\it Suzaku} observation (see Paper I).

\begin{figure}
\begin{center}
  \includegraphics[width=75mm]{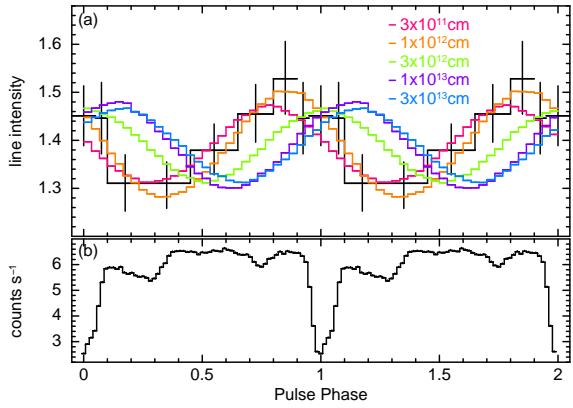}
\end{center}
\caption{
The intensity modulation of the iron line and the pulse shape of GX\,1+4 observed with {\it Suzaku} (see Paper I).
In panel (a), color lines show five samples of the simulated modulation curve.  
The angles,  $i$ and $\beta$, are tuned to have  the amplitude of 7~\%  and  the dip condition ($i=\beta$).
Assumed radii of circumstellar matter are indicated by different colors, red ($r=3\times10^{11}$~cm), orange ($r=1\times10^{12}$~cm), green ($3\times10^{12}$~cm), purple ($r=1\times10^{13}$~cm) and blue ($r=3\times10^{13}$~cm), respectively.
Intensities of the iron K$_{\alpha}$ line obtained from {\it Suzaku} observation are also plotted with black markers in the same panel.
Panel (b) shows the epoch folded light curve in a energy-range of 2--10~keV obtained from {\it Suzaku} observation.
}
\label{fig:gxeflc}
\end{figure}

The combination of the two angles, $i$ and $\beta$, can also be restricted by an interpretation of the pulse profile.
The pulse profile of GX\,1+4 in lower energy range (up to 10~keV) has a prominent sharp dip, as shown in Figure~\ref{fig:gxeflc}\,(b).
This sharp dip is interpreted as an obscuration or  eclipse of X-ray emitting region at a vicinity of a neutron star surface by the accretion column to the magnetic poles of the pulsar \citep{Galloway2000}.
Thus, the hot spot (i.e. magnetic pole) of the pulsar being at a base of the accretion column should face in the direction along the line of sight at least once a period.
Therefore, $i$ should be equal to $\beta$, with a certain acceptable range.
We estimated an angle of accretion column viewed from a magnetic pole from  the duration of the dip in the pulse profile.  
The dip was fitted by a Gaussian function and we got a width of the dip as $0.096\pm0.005$ (FWHM) in a unit of phase.
This value is equivalent to $\theta_{\rm FWHM}=34.6\pm0.2$~deg in angle.
We therefore estimated the acceptable range as $i-17^{\circ} \leq \beta \leq i+17^{\circ}$.
In Figure~\ref{fig:ampdist}, the green-solid lines indicate $i=\beta$ and the green-dotted lines indicate the acceptable range between $i$ and $\beta$.

The finite light speed effect constrains the modulation phase of the fluorescent line with regard to the pulse shape of the continuum emission.
In the case of GX\,1+4, the sharp dip is considered to occur when a direct photon from a magnetic pole is detected by the observer.
In Figure~\ref{fig:gxeflc}, intensity-modulation of iron K$_{\alpha}$ emission line of GX\,1+4 obtained from {\it Suzaku} and five typical calculated modulation samples, with amplitude of approximately 7~\% and satisfy $i=\beta$, are plotted as a function of the pulse phase.
The simulated modulation curve with $r=1\times10^{12}$~cm can reproduce the data modulation reasonably, 
but those with $r\ge3\times10^{12}$~cm mismatch the data plots because of their inconsistent phase shifts.
On the other hands, the modulation with amplitude of 7~\% can not be reproduced with $r\le1\times10^{11}$~cm (see  Figure~\ref{fig:ampdist}).

In conclusion, even if the fluorescent matter is homogenous and uniform, the observed intensity modulation can be explained by the matter extending up to $1\times10^{12}$~cm from the neutron star.
This distance is much larger than the co-rotation radius and the Alfv\'en radius, even for the extremely strong magnetic field of $10^{14}$~G. 

Here we discussed only one fluorescent region, though the emission from the neutron star arises from its two poles.
Several previous works suggest the existence of an accretion disk in GX~1+4.
\cite{ChakrabartyRoche1997} argued that  the optical emission lines suggests the existence of  thermal ultraviolet radiation from an accretion disk.
\cite{Rea2005} discussed reflection by a torus-like accretion disk.
\cite{Galloway2001} interpreted the dip in pulse profile as being due to the eclipse of the X-ray emitting region by the accretion curtain, which is formed by accretion flow along with magnetic field from inner edge of the accretion disk, showing a schematic figure of accretion geometry. 
The size of the accretion disk can be expected to be $1\times10^{12}$~cm from its orbital period  \citep{Pereira1999},  
and thus the fluorescent emission from the other side of the neutron star may be hidden by the accretion disk.  

\section{Discussion}
The finite light speed effect causes an intensity modulation  of fluorescent lines, with the rotation period of X-ray pulsars.
This is due to the apparent movement of the fluorescent region which is exposed to X-rays from a pulsar.
\edit3{The finite light speed effect thus can be expected to act across the X-ray pulsar source population.}
Therefore, for an interpretation of the intensity modulation of fluorescent lines from  X-ray pulsars, this effect should be taken into account.
This effect is efficient, if the size of the fluorescent matter is comparable to the pulse period multiplied by the speed of light or, in other words, the apparent movement speed of the fluorescent region is comparable to the speed of light.

In the case of GX\,1+4, the previously estimated fluorescent size ($\sim 10^{12}$ cm) by \citet{Kotani1999}, if it is true, matches to the above condition
 and  relatively strong effect can be expected.
Actually, the intensity modulation was detected and could be explained by this finite light speed effect.

However, in practice, the matter is not spherically symmetric and there should be various causes of the time variation of the fluorescent line.
\edit3{In the case of accretion powered X-ray pulsars, its circumstellar condition will change as function of its binary orbital phase.}
Several authors have reported the variations of the emission line parameters depending on the orbital phase of the binary system contained X-ray pulsar, such as Centaurus\,X-3, GX\,301-2 and Vela\,X-1 (\citealt{Ebisawa:1996,Naik:2011aa,Endo2002,Odaka2013}).
On the other hand the modulation due to the finite light speed effect should not change with the orbital phase.
Therefore,  the observation of the spin phase modulation at the various orbital phase would help to distinguish them.
   
It might be possible that the emission lines from photo-ionized plasma, exposed by a variable X-ray sources, also show similar effect by the finite speed of light.
In this case we have to consider the smearing effect by a  re-combination time scale as well as the light crossing time of the emission region.
Also even in pulsars with a short spin period  such as Hercules\,X-1 and 4U\,1626-67, 
if the emission region is comparable to the pulse period multiplied by the speed of light and if the line photons are created by the fluorescence or by the photo-ionization, 
we have to examine the finite light speed effect for the discussion of the line flux modulation.

\acknowledgments
This work was  partially supported by the Ministry of Education, Culture, Sports, 
Science and Technology (MEXT), Grant-in-Aid for Science Research 25400237, and the 
MEXT Supported Program for the Strategic Research Foundation at Private Universities, 
2014-2018. This research was carried out by using data obtained from the Data Archive 
and Transmission System (DARTS), provided by Center for Science-satellite Operation and 
Data Archive (C-SODA) at ISAS/JAXA.






\begin{thebibliography}{}
\expandafter\ifx\csname natexlab\endcsname\relax\def\natexlab#1{#1}\fi

\bibitem[{{Angelini} {et~al.}(1995){Angelini}, {White}, {Nagase}, {Kallman},
  {Yoshida}, {Takeshima}, {Becker}, \& {Paerels}}]{Angelini1995}
{Angelini}, L., {White}, N.~E., {Nagase}, F., {et~al.} 1995, \apjl, 449, L41

\bibitem[{{Basko}(1980)}]{Basko1980}
{Basko}, M.~M. 1980, \aap, 87, 330

\bibitem[{{Beri} {et~al.}(2015){Beri}, {Paul}, \& {Dewangan}}]{Beri2015}
{Beri}, A., {Paul}, B., \& {Dewangan}, G.~C. 2015, \mnras, 451, 508

\bibitem[{{Chakrabarty} \& {Roche}(1997)}]{ChakrabartyRoche1997}
{Chakrabarty}, D., \& {Roche}, P. 1997, \apj, 489, 254

\bibitem[{{Choi} {et~al.}(1994){Choi}, {Nagase}, {Makino}, {Dotani},
  {Kitamoto}, \& {Takahama}}]{Choi1994}
{Choi}, C.~S., {Nagase}, F., {Makino}, F., {et~al.} 1994, \apj, 437, 449

\bibitem[{{Ebisawa} {et~al.}(1996){Ebisawa}, {Day}, {Kallman}, {Nagase},
  {Kotani}, {Kawashima}, {Kitamoto}, \& {Woo}}]{Ebisawa:1996}
{Ebisawa}, K., {Day}, C.~S.~R., {Kallman}, T.~R., {et~al.} 1996, \pasj, 48, 425

\bibitem[{{Endo} {et~al.}(2002){Endo}, {Ishida}, {Masai}, {Kunieda}, {Inoue},
  \& {Nagase}}]{Endo2002}
{Endo}, T., {Ishida}, M., {Masai}, K., {et~al.} 2002, \apj, 574, 879

\bibitem[{{F{\"u}rst} {et~al.}(2011){F{\"u}rst}, {Suchy}, {Kreykenbohm},
  {Barrag{\'a}n}, {Wilms}, {Pottschmidt}, {Caballero}, {Kretschmar},
  {Ferrigno}, \& {Rothschild}}]{Furst2011}
{F{\"u}rst}, F., {Suchy}, S., {Kreykenbohm}, I., {et~al.} 2011, \aap, 535, A9

\bibitem[{{Galloway} {et~al.}(2000){Galloway}, {Giles}, {Greenhill}, \&
  {Storey}}]{Galloway2000}
{Galloway}, D.~K., {Giles}, A.~B., {Greenhill}, J.~G., \& {Storey}, M.~C. 2000,
  \mnras, 311, 755

\bibitem[{{Galloway} {et~al.}(2001){Galloway}, {Giles}, {Wu}, \&
  {Greenhill}}]{Galloway2001}
{Galloway}, D.~K., {Giles}, A.~B., {Wu}, K., \& {Greenhill}, J.~G. 2001,
  \mnras, 325, 419

\bibitem[{{Kohmura} {et~al.}(2001){Kohmura}, {Kitamoto}, \&
  {Torii}}]{Kohmura2001}
{Kohmura}, T., {Kitamoto}, S., \& {Torii}, K. 2001, \apj, 562, 943

\bibitem[{{Kotani} {et~al.}(1999){Kotani}, {Dotani}, {Nagase}, {Greenhill},
  {Pravdo}, \& {Angelini}}]{Kotani1999}
{Kotani}, T., {Dotani}, T., {Nagase}, F., {et~al.} 1999, \apj, 510, 369

\bibitem[{{Naik} {et~al.}(2011){Naik}, {Paul}, \& {Ali}}]{Naik:2011aa}
{Naik}, S., {Paul}, B., \& {Ali}, Z. 2011, \apj, 737, 79

\bibitem[{{Odaka} {et~al.}(2013){Odaka}, {Khangulyan}, {Tanaka}, {Watanabe},
  {Takahashi}, \& {Makishima}}]{Odaka2013}
{Odaka}, H., {Khangulyan}, D., {Tanaka}, Y.~T., {et~al.} 2013, \apj, 767, 70

\bibitem[{{Pereira} {et~al.}(1999){Pereira}, {Braga}, \&
  {Jablonski}}]{Pereira1999}
{Pereira}, M.~G., {Braga}, J., \& {Jablonski}, F. 1999, \apjl, 526, L105

\bibitem[{{Rea} {et~al.}(2005){Rea}, {Stella}, {Israel}, {Matt}, {Zane},
  {Segreto}, {Oosterbroek}, \& {Orlandini}}]{Rea2005}
{Rea}, N., {Stella}, L., {Israel}, G.~L., {et~al.} 2005, \mnras, 364, 1229

\bibitem[{{Sako} {et~al.}(2002){Sako}, {Kahn}, {Paerels}, {Liedahl},
  {Watanabe}, {Nagase}, \& {Takahashi}}]{Sako2002}
{Sako}, M., {Kahn}, S.~M., {Paerels}, F., {et~al.} 2002, in High Resolution
  X-ray Spectroscopy with XMM-Newton and Chandra, ed. G.~{Branduardi-Raymont}

\bibitem[{{Schulz} {et~al.}(2001){Schulz}, {Chakrabarty}, {Marshall},
  {Canizares}, {Lee}, \& {Houck}}]{Schulz2001}
{Schulz}, N.~S., {Chakrabarty}, D., {Marshall}, H.~L., {et~al.} 2001, \apj,
  563, 941

\bibitem[{{Suchy} {et~al.}(2012){Suchy}, {F{\"u}rst}, {Pottschmidt},
  {Caballero}, {Kreykenbohm}, {Wilms}, {Markowitz}, \&
  {Rothschild}}]{Suchy2012}
{Suchy}, S., {F{\"u}rst}, F., {Pottschmidt}, K., {et~al.} 2012, \apj, 745, 124

\bibitem[{{Vasco} {et~al.}(2013){Vasco}, {Staubert}, {Klochkov}, {Santangelo},
  {Shakura}, \& {Postnov}}]{Vasco2013}
{Vasco}, D., {Staubert}, R., {Klochkov}, D., {et~al.} 2013, \aap, 550, A111

\bibitem[{{Vrtilek} {et~al.}(2005){Vrtilek}, {Raymond}, {Boroson}, \&
  {McCray}}]{Vrtilek2005}
{Vrtilek}, S.~D., {Raymond}, J.~C., {Boroson}, B., \& {McCray}, R. 2005, \apj,
  626, 307

\bibitem[{{Wojdowski} {et~al.}(2003){Wojdowski}, {Liedahl}, {Sako}, {Kahn}, \&
  {Paerels}}]{Wojdowski2003}
{Wojdowski}, P.~S., {Liedahl}, D.~A., {Sako}, M., {Kahn}, S.~M., \& {Paerels},
  F. 2003, \apj, 582, 959

\bibitem[{Yoshida {et~al.}(2017)Yoshida, Kitamoto, Suzuki, Hoshino, Naik, \&
  Jaisawal}]{Yoshida2017}
Yoshida, Y., Kitamoto, S., Suzuki, H., {et~al.} 2017, The Astrophysical
  Journal, 838, 30

\bibitem[{{Zane} {et~al.}(2004){Zane}, {Ramsay}, {Jimenez-Garate}, {Willem den
  Herder}, \& {Hailey}}]{Zane2004}
{Zane}, S., {Ramsay}, G., {Jimenez-Garate}, M.~A., {Willem den Herder}, J., \&
  {Hailey}, C.~J. 2004, \mnras, 350, 506

\end{thebibliography}

%


\listofchanges

\end{document}